\pdfoutput=1
\RequirePackage{fix-cm}
\documentclass[british]{article}

\usepackage{spconf} % Conference template

\usepackage[utf8]{inputenc}
\usepackage[T1]{fontenc}
\usepackage[british]{babel}

\usepackage{pifont}

\usepackage{graphicx}
\usepackage{xcolor}
\graphicspath{{./figures/}}
\DeclareGraphicsExtensions{.pdf,.png,.jpeg,.jpg}

\usepackage{amsmath} % cmex10 ensures that only type 1 fonts are used at all font sizes
\usepackage{amssymb}
\usepackage{bm} % Bold maths
\usepackage{esint} % More integrals for computer modern math fonts
\usepackage{commath} % Provides \dif upright d
\usepackage{nicefrac} % Provides \nicefrac

\usepackage{subcaption} % Recommended, e.g., with the latest ACM template

\usepackage{booktabs} % Better-looking, more formal tables
\usepackage{array} % Patches and improves the array and tabular environments

\usepackage{multirow} % Enables table cells to span more than one row
\usepackage{rotating} % Enables rotating text in tables

\usepackage{enumitem} % Better control of lists and their spacing
\setlist{nolistsep} % Removes vertical spacing in lists

\usepackage[unicode=true,pdfusetitle,%
 pdfauthor={Shivam Mehta, Ruibo Tu, Jonas Beskow, Éva Székely, Gustav Eje Henter},%
 pdfkeywords={diffusion models, flow matching, speech synthesis, text-to-speech, acoustic modelling},%
 bookmarks=true,bookmarksnumbered=true,bookmarksopen=false,%
 breaklinks=true,pdfborder={0 0 0},backref=false,colorlinks=true,citecolor=blue]{hyperref}

\usepackage[capitalise]{cleveref} % Use \cref to automatically include, e.g., "Sec." in \ref (must be loaded after hyperref)
\Crefname{section}{Section}{Sections}
\crefname{section}{Sec.}{Secs.}
\Crefname{figure}{Figure}{Figures}
\crefname{figure}{Fig.}{Figs.}
\Crefname{table}{Table}{Tables}
\crefname{table}{Table}{Tabs.}

\renewcommand{\mid}{\,\ifnum\currentgrouptype=16 \middle\fi|\,}

\newcommand{\real}{\mathbb{R}}

\newcommand{\x}{\boldsymbol{x}}

\newcommand{\customurl}[1]{\url{#1}} % Monospaced URLs
\newcommand{\webpageurl}{https://shivammehta25.github.io/Matcha-TTS/}
\newcommand{\webpageurltext}{shivammehta25.github.io/Matcha-TTS/}

\ninept % Use minimum tolerated font size
\newcommand{\tablebf}[1]{%
\pdfliteral direct {2 Tr 0.5 w}% The second factor is the boldness
#1%
\pdfliteral direct {0 Tr 0 w}%
}

\let\oldmarginpar\marginpar
\renewcommand\marginpar[1]{\-\oldmarginpar[\raggedleft\footnotesize #1]%
{\raggedright\footnotesize #1}}

\title{Matcha-TTS: A fast TTS architecture with conditional flow matching}
\name{Shivam Mehta, Ruibo Tu, Jonas Beskow, {\'E}va Sz{\'e}kely, Gustav Eje Henter\thanks{This work was partially supported by the Wallenberg AI, Autonomous Systems and Software Program (WASP) funded by the Knut and Alice Wallenberg Foundation
and by the Industrial Strategic Technology Development Program (grant no.\ 20023495) funded by MOTIE, Korea.}}
\address{Division of Speech, Music and Hearing, KTH Royal Institute of Technology, Stockholm, Sweden}
\begin{document}
\maketitle
\begin{abstract}
%The abstract should contain about 100 to 150 words
%The online interface says "Limit the abstract text to a maximum of 200 words."
%Abstract goes here.
We introduce Matcha-TTS, a new encoder-decoder architecture for speedy TTS acoustic modelling, trained using optimal-transport conditional flow matching (OT-CFM). This yields an ODE-based decoder capable of high output quality in fewer synthesis steps than models trained using score matching. Careful design choices additionally ensure each synthesis step is fast to run. The method is probabilistic, non-autoregressive, and learns to speak from scratch without external alignments. Compared to strong pre-trained baseline models, the Matcha-TTS system has the smallest memory footprint, rivals the speed of the fastest model on long utterances, and attains the highest mean opinion score in a listening test.
%Diffusion-based generative models are producing excellent quality in, e.g., TTS, but generally require many steps for accurate output generation, slowing down synthesis. We introduce Matcha-TTS, a new encoder-decoder architecture for speedy TTS acoustic modelling. Our use of optimal-transport conditional flow matching (OT-CFM) for training yields high quality in fewer synthesis steps compared to TTS based on score-matching, whilst careful design choices also ensure each synthesis step is fast to run. The method is non-autoregressive, generalises well to long utterances, and learns to speak from scratch without external alignments. Compared to strong pre-trained baseline models, the Matcha-TTS system has the smallest footprint, rivals the speed of the fastest models on long utterances, and attains the highest mean opinion score in a listening test.
\end{abstract}
\section{Introduction}
\label{sec:intro}
Diffusion probabilistic models (DPMs) (cf.\ \cite{song2019generative}) are currently setting new standards in deep generative modelling on continuous-valued data-generation tasks such as image synthesis \cite{dhariwal2021diffusion,rombach2022high}, motion synthesis \cite{alexanderson2023listen,mehta2023diff}, and speech synthesis \cite{chen2021wavegrad, chen2021wavegrad2, popov2021grad, jeong2021diff, kongdiffwave} -- the topic of this paper.
%\cite{dhariwal2021diffusion, rombach2022high, song2019generative, nichol2021improved, nichol2022glide, saharia2022photorealistic}, motion synthesis \cite{ao2023gesturediffuclip, mehta2023diff, alexanderson2023listen, zhang2023diffmotion}.
% This includes TTS [cite examples]
%and text to speech (TTS) is no exception, various state-of-the-art approaches employ diffusion models to synthesise speech \cite{popov2021grad, chen2021wavegrad, chen2021wavegrad2, kongdiffwave, jeong2021diff}.
% Diffusion-based models represent probability distributions as differential equations that transform noise into samples from the desired target distribution. This is similar to continuous normalising flows, but diffusion models are faster to train thanks to techniques such as score matching
DPMs define a diffusion process which transforms the \emph{data} (a.k.a.\ \emph{target}) distribution to a \emph{prior} (a.k.a.\ \emph{source}) distribution, e.g., a Gaussian.
They then learn a \emph{sampling process} that reverses the diffusion process.
%and transforms the source distribution into the data distribution.
The two processes can be formulated as forward- and reverse-time stochastic differential equations (SDEs) \cite{song2021score}.
Solving a reverse-time SDE initial value problem  generates samples from the learnt data distribution.
Furthermore, each reverse-time SDE has a corresponding ordinary differential equation (ODE), called the \emph{probability flow ODE} \cite{song2021score,albergo2022building}, which describes (and samples from) the exact same distribution as the SDE.
%sharing the same marginal distribution at each time step, which is . 
The probability flow ODE is a deterministic process for turning source samples into data samples,
%using the score functions \cite{song2019generative} for synthesis and is much similar to
similar to continuous-time normalising flows (CNF) \cite{chen2018neural}, but without the need to backpropagate through expensive ODE solvers or approximate the reverse ODE using adjoint variables \cite{chen2018neural}.
%\cite{grathwohl2019ffjord} or to approximate the reverse ODE using adjoint variables \cite{chen2018neural}.

The SDE formulation of DPMs is trained by approximating the score function (the gradients of the log probability density) of the data distribution \cite{song2021score}.
The training objective takes the form of a mean squared error (MSE) which can be derived from an evidence lower bound (ELBO) on the likelihood.
This is fast and simple and, unlike typical normalising flow models, does not impose any restrictions on model architecture.
But whilst they allow efficient training without numerical SDE/ODE solvers, DPMs suffer from slow synthesis speed, since each sample requires numerous iterations (steps), computed in sequence, to accurately solve the SDE.
Each such step requires that an entire neural network be evaluated.
This slow synthesis speed has long been the main practical issue with DPMs.

This paper introduces \emph{Matcha-TTS}\footnote{We call our approach Matcha-TTS because it uses flow matching for TTS, and because the name sounds similar to ``matcha tea'', which some people prefer over Taco(tron)s.}, a probabilistic and non-autoregressive, fast-to-sample-from TTS acoustic model based on continuous normalising flows.
%that jointly learns to speak and align.
There are two main innovations:
\begin{enumerate}
\item To begin with, we propose an improved encoder-decoder TTS architecture that uses a combination of 1D CNNs and Transformers in the decoder. This reduces memory consumption and is fast to evaluate, improving synthesis speed.
%In Matcha-TTS, we propose a novel encoder-decoder TTS architecture employing a combination of 1D convolutional and transformers in the decoder. This aims to enhance the synthesising speed and memory consumption.
\item Second, we train these models using optimal-transport conditional flow matching (OT-CFM) \cite{lipman2023flow},
%recently proposed by Lipman et al.\ \cite{lipman2023flow}.
which is a new method to learn ODEs that sample from a data distribution.
Compared to conventional CNFs and score-matching probability flow ODEs, OT-CFM defines simpler paths from source to target, enabling accurate synthesis in fewer steps than DPMs.
\end{enumerate}
%Further, we employ the recently introduced framework of conditional flow matching for training and sampling from the complex data distribution. More specifically, for training Matcha-TTS we utilise the optimal transport conditional flow matching (OT-CFM) formulation proposed in \cite{lipman2023flow}. These OT-CFM paths are simpler than the paths traversed by DPMs. We show that these simpler paths result in faster generation and better performance with less number of iterations as compared to previous score-matching approaches.
Experimental results demonstrate that both innovations accelerate synthesis, reducing the trade-off between speed and synthesis quality.
Despite being fast and lightweight, Matcha-TTS learns to speak and align without requiring an external aligner.
Compared to strong pre-trained baseline models, Matcha-TTS achieves fast synthesis with better naturalness ratings.
% achieve similar speech naturalness with faster synthesis or better speech naturalness with similar synthesis or both synthesis qualities.
Audio examples and code are provided at \href{\webpageurl}{\webpageurl}.

\section{Background}
\label{sec:background}
%In this section, we review prior work in encoder-decoder acoustic modelling for TTS, including DPM-based methods and how their training can be improved.
%, and contrast these against the proposed model.
\subsection{Recent encoder-decoder TTS architectures}

DPMs have been applied to numerous speech-synthesis tasks with impressive results, including waveform generation \cite{chen2021wavegrad,kongdiffwave} and end-to-end TTS \cite{chen2021wavegrad2}.
%Wavegrad \cite{chen2021wavegrad} and DiffWave \cite{kongdiffwave} were amongst the first to use them for waveform synthesis. Subsequently, they have also been used for end-to-end synthesis in WaveGrad 2\cite{chen2021wavegrad2}.
Diff-TTS \cite{jeong2021diff} was first to apply DPMs for acoustic modelling.
Shortly after, Grad-TTS \cite{popov2021grad} conceptualised the diffusion process as an SDE.
%, a formulation used in subsequent models, e.g., \cite{vovk2022fast}.
%This was swiftly adopted by other models like Grad-StyleSpeech \cite{kang2023gradstylespeech} and Emo-Diff \cite{guo2023emodiff}, both performing few-/one-shot synthesis using an additional reference encoder.
Although these models, and descendants like Fast Grad-TTS \cite{vovk2022fast}, are non-autoregressive, \mbox{TorToiSe} \cite{betker2023better} demonstrated DPMs in an autoregressive TTS model with quantised latents.
%for transforming quantised latents from an autoregressive model into acoustic features.
%employed DPMs in a quantised latent domain where an autoregressive transformer first learned a series of quantised vectors and then transformed them into a spectrogram using DPMs.
%TorToiSe \cite{betker2023better} employed DPMs in a quantised latent domain where an autoregressive transformer first learned a series of quantised vectors and then transformed them into a spectrogram using DPMs.  

The above models -- like many modern TTS acoustic models -- use an encoder-decoder architecture with Transformer blocks
%\cite{vaswani2017attention}
in the encoder.
%An established architecture for encoding linguistic information is encoder transformer blocks \cite{vaswani2017attention}. Many modern and widely used architectures \cite{kim2020glow, kim2021vits, popov2021grad, ren2019fastspeech, lancucki2021fastpitch, ren2021fastspeech2} use transformer encoder architecture for their speed and the ability to capture long-range dependencies.
Many models, e.g., FastSpeech 1 and 2 \cite{ren2019fastspeech,ren2021fastspeech2},
%and FastPitch \cite{lancucki2021fastpitch},
use sinusoidal position embeddings for positional dependences.
This has been found to generalise poorly to long sequences; cf.\ \cite{press2022train}.
%failing to generalise for longer utterances,
Glow-TTS \cite{kim2020glow}, VITS \cite{kim2021vits}, and Grad-TTS instead use relative positional embeddings \cite{shaw2018self}.
%for improved generalisation over longer utterances.
Unfortunately, these
%do not differentiate between inputs beyond a threshold window length,
treat inputs outside a short context window as a ``bag of words'',
often resulting in unnatural prosody.
LinearSpeech \cite{zhang2021linearspeech} instead employed rotational position embeddings (RoPE) \cite{su2021roformer}, which have computational and memory advantages over relative embeddings and generalise to longer distances \cite{wennberg2021case,press2022train}.
Matcha-TTS thus uses Transformers with RoPE
%for positional dependence
in the encoder, reducing RAM use compared to Grad-TTS.
%However, these have not yet been used in ODE-based TTS architecture to the best of our knowledge.
We believe ours is the first SDE or ODE-based TTS method to use RoPE.
%employed RoPE for encoding positional information in a text encoder.

Modern TTS architectures also differ in terms of decoder network design.
The normalising-flow based methods Glow-TTS \cite{kim2020glow} and OverFlow \cite{mehta2023overflow} use dilated 1D-convolutions.
DPM-based methods like \cite{jeong2021diff,huang2022prodiff}
%, and WaveGrad \cite{chen2021wavegrad}
likewise use 1D convolutions to synthesise mel spectrograms.
%All these design choices as translation invariant along the time.
%For probabilistic acoustic feature generators such as Glow-TTS \cite{kim2020glow}, and OverFlow \cite{mehta2023overflow} use normalizing flows and employ invertible neural networks consisting of dilated 1D-convolutions. Other diffusion-based methods like Diff-TTS \cite{jeong2021diff}, ProDiff \cite{huang2022prodiff}, and \cite{chen2021wavegrad}, use 1D convolutions to synthesise mel spectrograms and treat them as translation invariant along the time. 
Grad-TTS \cite{popov2021grad}, in contrast, 
%However, Grad-TTS \cite{popov2021grad}
uses a U-Net with 2D-convolutions.
This treats mel spectrograms as images and implicitly assumes translation invariance in both time and frequency.
However, speech mel-spectra are not fully translation-invariant along the frequency axis, and 2D decoders generally require more memory as they introduce an extra dimension to the tensors.
%with extra dimensions.
Meanwhile, non-probabilistic models like FastSpeech 1 and 2
%and FastPitch
have demonstrated that decoders with (1D) Transformers can learn long-range dependencies and fast, parallel synthesis.
%For non-probabilistic acoustic feature generators like  exhibit the advantage of using transformers for learning long-range dependencies and parallel synthesis.
Matcha-TTS also uses Transformers in the decoder, but in a 1D U-Net design inspired by the 2D U-Nets in the Stable Diffusion image-generation model \cite{rombach2022high}.

Whilst some TTS systems, e.g., FastSpeech \cite{ren2019fastspeech}, rely on externally-supplied alignments, most systems are capable of learning to speak and align at the same time,
%using only parallel audio and text prompts.
although it has been found to be important to encourage or enforce monotonic alignments
%\cite{tachibana2018efficiently,watts2019where,mehta2022neuralhmm}
\cite{watts2019where,mehta2022neuralhmm}
for fast and effective training.
One mechanism for this is monotonic alignment search (MAS), used by, e.g., Glow-TTS \cite{kim2020glow} and VITS \cite{kim2021vits}.
%, and FastPitch 1.1 \cite{lancucki2021fastpitch}.
%and RAD-TTS++ \cite{valle2023highacoustic}.
Grad-TTS \cite{popov2021grad}, in particular, uses a MAS-based mechanism
%\cite{badlani2022one}
which they term \emph{prior loss} to quickly learn to align input symbols with output frames.
These alignments are also used to train a deterministic duration predictor minimising MSE in the log domain.
Matcha-TTS uses these same methods for alignment and duration modelling.
Finally, Matcha-TTS differs by using \emph{snake beta} activations from BigVGAN \cite{lee2023bigvgan}
%, a modification of the snake (sinusiodal) activations from \cite{ziyin2020neural},
in all decoder feedforward layers.

\subsection{Flow matching and TTS}
%\begin{itemize}
%\item Currently, the strongest TTS systems generally use diffusion models [citations]
%\item Reference the flow matching paper and state the advantages of the technique
%\item Reference the conditional flow matching paper and state what it adds
%\item State applications where flow matching has been used, with citations
%\item State that, so far, flow matching and conditional flow matching have only been used once for TTS, namely VoiceBox
%\item Describe VoiceBox in one or two sentences
%\item Describe the most important differences between our work and VoiceBox
%\end{itemize}

Currently, some of the highest-quality TTS systems either utilise DPMs
%\cite{popov2021grad, shen2023naturalspeech, betker2023better}
\cite{popov2021grad, betker2023better}
or discrete-time normalising flows \cite{kim2021vits,mehta2023overflow}, with continuous-time flows being less explored.
Lipman et al.\ \cite{lipman2023flow} recently introduced a framework for synthesis using ODEs that unifies and extends probability flow ODEs and CNFs.
They were then able to present an efficient approach to learn ODEs for synthesis, using a simple vector-field regression loss called \emph{conditional flow matching} (CFM), as an alternative to learning score functions for DPMs or using numerical ODE solvers at training time like classic CNFs \cite{chen2018neural}.
Crucially, by leveraging ideas from optimal transport, CFM can be set up to yield ODEs that have simple vector fields that change little during the process of mapping samples from the source distribution onto the data distribution, since it essentially just transports probability mass along straight lines.
This technique is called \emph{OT-CFM};
\emph{rectified flows} \cite{liu2023flow} represent concurrent work with a similar idea.
The simple paths mean that the ODE can be solved accurately using few discretisation steps, i.e., accurate model samples can be drawn with fewer neural-network evaluations than DPMs, enabling much faster synthesis for the same quality.

CFM is a new technique that differs from earlier approaches to speed up SDE/ODE-based TTS, which most often were based on distillation (e.g., \cite{huang2022prodiff,vovk2022fast,ye2023comospeech}).
Prior to Matcha-TTS, the only public preprint on CFM-based acoustic modelling was the Voicebox model from Meta \cite{le2023voicebox}.
%VoiceBox \cite{le2023voicebox} is the only application that has employed CFMs in speech synthesis.
Voicebox (VB) is a system that performs various text-guided speech-infilling tasks based on large-scale training data, with its English variant (VB-En) being trained on 60k hours of proprietary data.
%and has a staggering 330M parameters.
VB differs substantially from Matcha-TTS:
VB performs TTS, denoising, and text-guided acoustic infilling trained using a combination of masking and CFM, whereas Matcha-TTS is a pure TTS model trained solely using OT-CFM.
VB uses convolutional positional encoding with AliBi \cite{press2022train} self-attention bias, whilst our text encoder uses RoPE.
In contrast to VB, we train on standard data and make code and checkpoints publicly available.
%, whilst VB is proprietary.
VB-En consumes 330M parameters, which is 18 times larger than the Matcha-TTS model in our experiments.
Also, VB uses external alignments for training whereas Matcha-TTS learns to speak without them.

%; however, it uses an external aligner i.e. Montreal Force Aligner (MFA) \cite{mcauliffe2017_montreal} and does not jointly learn to align and generate spectrogram. Another drawback is the unavailability of source code for VB, and due to its staggering size, generating speech from it without access to powerful GPU clusters is challenging and expensive. Our work differs from VB in many ways. Firstly, the primary training objective of VB was to infill speech given audio context and text using a masked language model training regime with the assistance of CFM, while we train a text-to-speech system with CFM. Secondly, VB uses convolutional positional encoding with AliBi self-attention bias, while for text encoding we use much faster and simpler RoPE embeddings. Most importantly, VB requires an external aligner to train while the proposed method learns to align from the data itself without the need for an external aligner. Practically, The duration model in VB is a transformer with 28M parameters in its smallest configuration, while the size of our entire architecture is 18M. This further demonstrates the effectiveness of our method having \textbf{18x} less number of parameters and making it practical to train a TTS system in an ordinary setting.

\section{Method}
\label{sec:method}
We now outline flow-matching training (in \cref{ssec:cfm}) and then (in \cref{ssec:matcha}) give details on our proposed TTS architecture.

\subsection{Optimal-transport conditional flow matching}
\label{ssec:cfm}
%\begin{itemize}
%\item Briefly (maximum half a page) give a technical description of conditional flow matching
%\item 1. what is the probability path from p0 to p1.
%\item 2. The Flow Matching objective is to match such probability paths. 
%\item 3. why and what is Conditional flow matching
%\item 4. introduce the OT-CFM used in this work.
%\item Only include maths if it helps tell the story better/more clearly
%\end{itemize}
We here give a high-level overview of flow matching, first introducing the probability-density path generated by a vector field and then
%flow matching and CFM objectives.
leading into the OT-CFM objective used in our proposed method.
Notation and definitions mainly follow \cite{lipman2023flow}.

Let 
% $\real^d$ denote the dataspace with data points 
$\x$ denote an observation in the data space $\real^d$, sampled from a complicated, unknown data distribution $q(\x)$. 
A \emph{probability density path} is a time-dependent probability density function, $p_t: [0,1]\times \real^d \rightarrow \real>0$.
One way to generate samples from the data distribution $q$ is to construct a probability density path $p_t$, where $t \in [0,1]$ and $p_0(\x) = \mathcal{N}(\x; \boldsymbol{0},\boldsymbol{I})$ is a prior distribution, such that $p_1(\x)$ approximates the data distribution $q(\x)$.
For example, CNFs first define a vector field $\boldsymbol{v}_t: [0, 1] \times \real^d \rightarrow \real^d$, which generates the flow $\phi_t: [0, 1] \times \real^d \rightarrow \real^d$ through the ODE 
\begin{align}
\tfrac{d}{dt}\phi_t(\x)
& = \boldsymbol{v}_t(\phi_t(\x))
\text{;}
\qquad
\phi_0(\x)
= \x
\text{.}
\label{eq:ode}
\end{align}
This generates the path $p_t$ as the marginal probability distribution of the data points.
We can sample from the approximated data distribution $p_1$ by solving the initial value problem in Eq.\ \eqref{eq:ode}. 
%of such ODEs.

% To learn a transformation from a simple prior distribution $p_0(\x) = \mathcal{N}(\x; \mathbf{0},\boldsymbol{I})$ to the data distribution $q$ we can employ CNFs. Let $\boldsymbol{v}_t: [0, 1] \times \real^d \rightarrow \real^d$  be a time dependent vector field that pushes the point from the prior $p_0$ to $q$ with the help of flow: $\phi_t: [0, 1] \times \real^d \rightarrow \real^d$ such that the flow can be defined with an ODE as 
% $$ \frac{d}{dt}\phi_t(x) = \boldsymbol{v}_t(\phi(x)) ; \ \ \ \  \phi_0(x) = x$$

Suppose there exists a known vector field $\boldsymbol{u}_t$ that generates a probability path $p_t$ from $p_0$ to $p_1\approx q$.
The flow matching loss is
\begin{align}
\mathcal{L}_{\mathrm{FM}}(\theta)
& = \mathbb{E}_{t, p_t(\x)}\Vert \boldsymbol{u}_t(\x) - \boldsymbol{v}_t(\x; \theta) \Vert^2
\text{,}
\end{align}
where $t\sim\mathbb{U}[0,1]$ and $\boldsymbol{v}_t(\x; \theta)$ is a neural network with parameters $\theta$.
Nevertheless, flow matching is intractable in practice because it is non-trivial to get access to the vector field $\boldsymbol{u}_t$ and the target probability $p_t$.
Therefore, conditional flow matching instead considers
\begin{align}
\mathcal{L}_{\mathrm{CFM}} (\theta)
& = \mathbb{E}_{t, q(\x_1),p_t(\x|\x_1)}\Vert \boldsymbol{u}_t(\x\vert \x_1) - \boldsymbol{v}_t(\x; \theta) \Vert^2
\text{.}
\end{align}
This replaces the intractable marginal probability densities and the vector field with conditional probability densities and conditional vector fields. Crucially, these are in general tractable and have closed-form solutions, and one can furthermore show that $\mathcal{L}_{\mathrm{CFM}}(\theta)$ and $\mathcal{L}_{\mathrm{FM}}(\theta)$ both have identical gradients with respect to $\theta$ \cite{lipman2023flow}.
\begin{figure}[!t]
\centering
\includegraphics[width=\columnwidth]{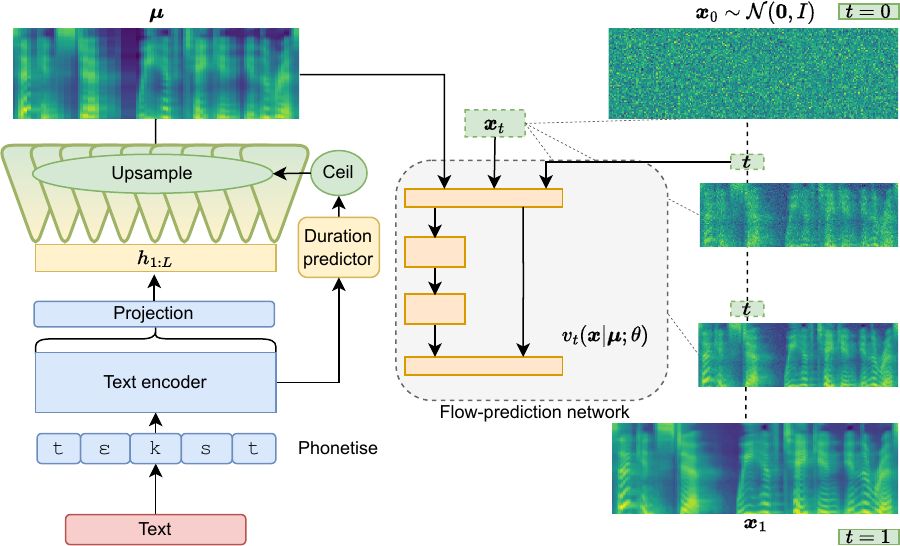}
\caption{Overview of the proposed approach at synthesis time.}
\label{fig: model architecture overview}
\vspace{-\baselineskip}
\end{figure}

Matcha-TTS is trained using optimal-transport conditional flow matching (OT-CFM) \cite{lipman2023flow}, which is a CFM variant with particularly simple gradients.
The OT-CFM loss function can be written
\begin{align}
%\resizebox{.99\hsize}{!}{$%
\mathcal{L} (\theta) & =
\mathbb{E}_{t, q(\x_1),p_0(\x_0)}
\Vert\boldsymbol{u}^{\mathrm{OT}}_t(\phi^{\mathrm{OT}}_t(\x)| \x_1)- \boldsymbol{v}_t(\phi^{\mathrm{OT}}_t(\x) | \boldsymbol{\mu}; \theta) \Vert^2
\text{,}
%$}
\end{align}
defining $\phi^{\mathrm{OT}}_t(\x) = (1 - (1-\sigma_{\mathrm{min}})t)\x_0 + t \x_1$ as the flow from $\x_0$ to $\x_1$ where each datum $\x_1$ is matched to a random sample $\x_0\sim\mathcal{N}(\boldsymbol{0},\boldsymbol{I})$ as in \cite{lipman2023flow}.
Its gradient vector field -- whose expected value is the target for the learning -- is then $\boldsymbol{u}^{\mathrm{OT}}_t(\phi^{\mathrm{OT}}_t(\x_0)\vert \x_1) = \x_1-(1-\sigma_{\mathrm{min}})\x_0$, which is linear, time-invariant, and only depends on $\x_0$ and $\x_1$.
These properties enable easier and faster training, faster generation, and better performance compared to DPMs.
%Which source and target points $\x_0$ and $\x_1$ that flow to each other is chosen using by solving an optimal transport problem (least total displacement).
%per minibatch.

In the case of Matcha-TTS, $\x_1$ are acoustic frames and $\boldsymbol{\mu}$ are the conditional mean values of those frames, predicted from text using the architecture described in the next section.
$\sigma_{\mathrm{min}}$ is a hyperparameter with a small value (\texttt{1e-4} in our experiments).

%Because of how simple the vector field $\boldsymbol{u}^{\mathrm{OT}}$ is, Matcha-TTS has much easier and faster training, faster generation, and better performance compared to diffusion models.

% Let $p_t$ be a probability path and $\boldsymbol{u}_t$ be the vector field corresponding to that path. We can parameterise the vector field $\boldsymbol{v}_t(x; \theta)$ by a neural network with parameters $\theta$ and this can be trained with a FM objective
% $$\mathcal{L}_{FM}(\theta) = \mathbb{E}_{t, p_t(x)}\Vert \boldsymbol{u}_t(x) - \boldsymbol{v}_t(x; \theta) \Vert^2$$
% where $t\sim\mathbb{U}[0,1]$ and $x\sim p_t(x)$.

% Conditional Flow Matching simplifies the Flow matching objective by replacing the untractable marginal distributions of vector fields with conditional distribution while generating identical gradients. 
% $$\mathcal{L}_{CFM} (\theta) = \mathbb{E}_{t, q(x_1),p_t(x|x_1)}\Vert \boldsymbol{u}_t(x\vert x_1) - \boldsymbol{v}_t(x; \theta) \Vert^2$$

\subsection{Proposed architecture}
\label{ssec:matcha}
%\begin{itemize}
%\item Give a technical description of the proposed TTS method
%\item First the new architecture
%\item The main thing that readers should take away is technical insight into why the relevant changes (rotary embeddings, 1D decoder with Transformers instead of 2D encoder with CNNs) reduce memory use and synthesis time
%\item Include an architecture plot
%\item Then describe how flow matching was applied to train the system. This can probably be very brief, given the work done by the previous subsection
%\item Things such as specific hyperparameters that represent an instantiation of the system are reserved for the experiments
%\item Name the approach; something like ``We call our approach \emph{Matcha-TTS} because it uses flow matching for TTS, and because the name sounds similar to ``matcha tea'', which some people prefer over Taco(tron)s.''
%\end{itemize}

Matcha-TTS is a non-autoregressive encoder-decoder architecture for neural TTS.
An overview of the architecture is provided in \cref{fig: model architecture overview}.
Text encoder and duration predictor architectures follow \cite{kim2020glow,popov2021grad}, but use rotational position embeddings \cite{su2021roformer} instead of relative ones.
Alignment and duration-model training follow use MAS and the prior loss $\mathcal{L}_{\mathrm{enc}}$ as described in \cite{popov2021grad}.
The predicted durations, rounded up, are used to upsample (duplicate) the vectors output by the encoder to obtain $\boldsymbol{\mu}$, the predicted average acoustic features (e.g., mel-spectrogram) given the text and the chosen durations.
This mean is used to condition the decoder that predicts the vector field $\boldsymbol{v}_t(\phi^{\mathrm{OT}}_t(\x_0) | \boldsymbol{\mu}; \theta)$ used for synthesis, but is not used as the mean for the initial noise samples $\x_0$ (unlike Grad-TTS).

\begin{figure}[!t]
\centering
\includegraphics[width=0.72\columnwidth]{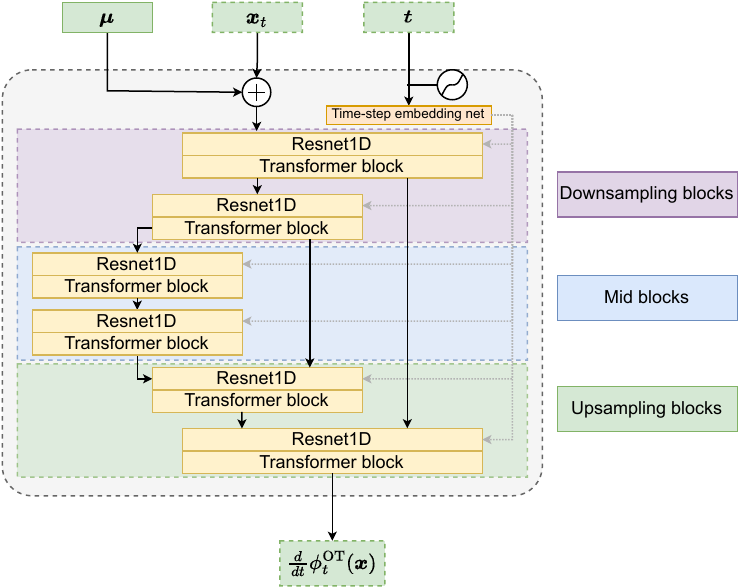}
\caption{Matcha-TTS decoder (the flow-prediction network in \cref{fig: model architecture overview}).}
\label{fig: decoder architecture}
\vspace{-\baselineskip}
\end{figure}

\cref{fig: decoder architecture} shows the Matcha-TTS decoder architecture.
Inspired by \cite{rombach2022high}, it is a U-Net containing 1D convolutional residual blocks to downsample and upsample the inputs,
with the flow-matching step $t\in[0,1]$ embedded as in \cite{popov2021grad}.
%based on an embedding of the flow matching step $t\in[0,1]$, in our experiments computed by a network comprising two affine layers stacked.
Each residual block is followed by a Transformer
%\cite{vaswani2017attention}
block, whose feedforward nets use snake beta activations \cite{lee2023bigvgan}.
These Transformers do not use any position embeddings, since between-phone positional information already has been baked in by the encoder, and the convolution and downsampling operations act to interpolate these between frames within the same phone and distinguish their relative positions from each other.
This decoder network is significantly faster to evaluate and consumes less memory than the 2D convolutional-only U-Net used by Grad-TTS \cite{popov2021grad}.

\section{Experiments}
\label{sec:experiments}
To validate the proposed approach we compared it to three pre-trained baselines in several experiments, including a listening test.
All experiments were performed on NVIDIA RTX 3090 GPUs.
See \href{\webpageurl}{\webpageurltext} for audio and code.

\subsection{Data and systems}
\label{ssec:systems}
%\begin{itemize}
%\item See our previous papers for how to write this
%\item Describe the data very briefly
%\item Introduce the three baseline systems (FastSpeech 2, Grad-TTS, VITS), give them a bold label, and motivate why each was included
%\item Emphasise that we are using strong pre-trained checkpoints in all cases
%\item Introduce the proposed system and its label
%\item Describe phonetisation and training (batch size, number of updates, GPUs) for the proposed system
%\item Describe that we also trained a version of Grad-TTS using CFM, to be able to separate the effect of the architectural changes we proposed from the effect of using CFM for training
%\item State that the diffusion-based systems can use a different number of function evaluations for numerically solving the differential equations at synthesis time, meaning that we can generate from them in multiple ways that allow trading off between speech quality and synthesis speed. State that we evaluate the proposed and baseline systems using several different NFEs to evaluate this trade-off
%\item Make sure that the difference between an artificial \emph{system} (FastSpeech 2, Grad-TTS, VITS, proposed, ablation) and a \emph{condition} (which can be multiple ways of synthesising from a system, and additionally includes VOC) is clear throughout
%\end{itemize}

We performed our experiments on the standard split of the LJ Speech dataset\footnote{\customurl{https://keithito.com/LJ-Speech-Dataset/}} (a female US English native speaker reading public-domain texts), training a version of the Matcha-TTS architecture on this data.
We used the same encoder and duration predictor (i.e., the same hyperparameters) as \cite{popov2021grad}, just different position embeddings in the encoder.
Our trained flow-prediction network (decoder) used two downsampling blocks, followed by two midblocks and two upsampling blocks, as shown in \cref{fig: decoder architecture}.
Each block had one Transformer layer with hidden dimensionality 256, 2 heads, attention dimensionality 64, and `snakebeta' activations \cite{lee2023bigvgan}.
Phonemizer\footnote{\customurl{https://github.com/bootphon/phonemizer}} \cite{bernard2021phonemizer} with the \texttt{espeak-ng} backend was used to convert input graphemes to IPA phones.
We trained for 500k updates on 2 GPUs with batch size 32 and learning rate \texttt{1e-4}, labelling our trained system \textbf{MAT}.

MAT was compared to three widely used neural TTS baseline approaches with pre-trained checkpoints available for LJ Speech, namely Grad-TTS\footnote{\customurl{https://github.com/huawei-noah/Speech-Backbones/tree/main/Grad-TTS}} \cite{popov2021grad} (label \textbf{GRAD}), a strong DPM-based acoustic model, FastSpeech 2 (\textbf{FS2}), a fast non-probabilistic acoustic model, and \textbf{VITS}\footnote{\customurl{https://github.com/jaywalnut310/vits}}, a strong probabilistic end-to-end TTS system with discrete-time normalising flows.
The baselines used the official checkpoints from the respective linked repositories.
For FS2, which does not provide an official implementation, we instead used the checkpoint from Meta's FairSeq\footnote{\customurl{https://github.com/facebookresearch/fairseq}}.
%implementation and pre-trained checkpoint available for LJ Speech .
To decouple the effects of CFM training from those due to the new architecture, we also trained the GRAD architecture using the OT-CFM objective instead, using the same optimiser hyperparameters as for MAT.
This produced the ablation labelled \textbf{GCFM}.
%To demonstrate the effectiveness of CFM and decouple its effect of architectural changes, we trained GRAD with CFM objective and named the condition \textbf{GCFM}.
For all acoustic models (i.e., all systems except VITS), we used the pre-trained HiFi-GAN \cite{kong2020hifi} LJ Speech checkpoint \texttt{LJ\_V1}\footnote{\customurl{https://github.com/jik876/hifi-gan/}} for waveform generation,
%containing 13.9M parameters followed by
with a denoising filter as introduced in \cite{prenger2019waveglow} at a strength of \texttt{2.5e-4}.
As a top line, our experiments also included vocoded held-out speech, labelled \textbf{VOC}.%
\begin{table}[!t]
\centering
\begin{tabular}{@{}l|cc|ccc@{}}
\toprule 
Condition & Params. & RAM & RTF ($\mu{\pm}\sigma$) & WER & MOS\tabularnewline
\midrule
VOC & 13.9M  & - & 0.001$\pm$0.001 & 1.97 & 4.13$\pm$0.09\tabularnewline
\midrule
FS2 & 41.2M  & \hphantom{0}6.0 & \tablebf{0.010}$\pm$0.004 & 4.18 & 3.29$\pm$0.09\tabularnewline
VITS & 36.3M & 12.4 & 0.074$\pm$0.083 & 2.52 & 3.71$\pm$0.08\tabularnewline
GRAD-10 & \tablebf{14.8M} & \hphantom{0}7.8 & 0.049$\pm$0.013 & 3.44 & 3.49$\pm$0.08\tabularnewline
GRAD-4 & \textquotedbl & \textquotedbl & 0.019$\pm$0.006 & 3.69 & 3.20$\pm$0.09\tabularnewline
GCFM-4 & \textquotedbl & \textquotedbl & 0.019$\pm$0.004 & 2.70 & 3.57$\pm$0.08\tabularnewline
\midrule
MAT-10 & 18.2M & \hphantom{0}\tablebf{4.8} & 0.038$\pm$0.019 & \tablebf{2.09} & \tablebf{3.84}$\pm$0.08\tabularnewline
MAT-4 & \textquotedbl & \textquotedbl & 0.019$\pm$0.008 & 2.15 & 3.77$\pm$0.07\tabularnewline
MAT-2 & \textquotedbl & \textquotedbl & 0.015$\pm$0.006 & 2.34 & 3.65$\pm$0.08\tabularnewline
\bottomrule
\end{tabular}
\caption{Conditions in the evaluation (with the NFE for ODE-based methods) and their number of parameters, minimum GPU RAM needed to train (GiB), real-time factor (including vocoding time) on the test set, ASR WER in percent, and mean opinion score with 95\%-confidence interval.
The best TTS condition in each column is bold.
%The vocoder (13.9M parameters) is in addition to all parameter counts except for VITS, which is end-to-end.
The parameter count and RTF for VOC pertain to the vocoder.
%VOC had a WER of 1.97 and a MOS of 4.13$\pm$0.07.
}
\label{tab:results}
\vspace{-\baselineskip}
\end{table}%

ODE-based models, e.g., DPMs, allow trading off speed against quality.
We therefore evaluated synthesis from the trained ODE-based systems with a different number of steps for the ODE solver.
Like \cite{popov2021grad}, we used the first-order Euler forward ODE-solver, where the number of steps is equal to the number of function (i.e., neural-network) evaluations, commonly abbreviated \emph{NFE}.
%(which determines how many times the decoder network must be evaluated).
This gave rise to multiple \emph{conditions} for some systems.
We labelled these conditions \textbf{MAT-$\boldsymbol{n}$}, \textbf{GRAD-$\boldsymbol{n}$}, and \textbf{GCFM-$\boldsymbol{n}$}, $n$ being the NFE.
%number of first-order Euler forward ODE-solver steps.
We used NFE 10 or less, since \cite{popov2021grad} reported that NFE 10 and 100 gave the same MOS for Grad-TTS (NFE 50 is the official code default).
All conditions used a temperature of 0.667 for synthesis, similar to \cite{popov2021grad}.
\cref{tab:results} provides an overview of the conditions in the evaluation.

%various conditions of proposed and baseline systems using several netwoek evaluations NFEs to evaluate this trade-off and computed their Real Time Factors (RTFs) with vocoder. The utterances were synthesised on \texttt{Nvidia RTX-3090} GPUs with batch size one using Euler numerical solver. 

\subsection{Evaluations, results, and discussion}
\label{ssec:results}
We evaluated our approach both objectively and subjectively.
First we measured parameter count and maximum memory use during training (at batch size 32 and fp16) of all systems, with results listed in \cref{tab:results}.
We see that MAT is approximately the same size as GRAD/GCFM, and smaller than all other systems.
In particular, it is smaller than VITS also after adding the vocoder (13.9M parameters) to the MAT parameter count.
More importantly, it uses less memory than all baselines, which (more than parameter count) is the main limiter on how large and powerful models that can be trained.

After training the systems, we assessed the synthesis speed and intelligibility of the different conditions, by computing the real time factor (RTF) mean and standard deviation when synthesising the test set, and evaluating the word error rate (WER) when applying the Whisper \texttt{medium}
%\footnote{\customurl{https://huggingface.co/openai/whisper-medium}}
\cite{radford2023robust} ASR system to the results, since the WERs of strong ASR systems correlate well with intelligibility \cite{taylor2021confidence}.
The results, in \cref{tab:results}, suggest that MAT is the most intelligible system, even using only two synthesis steps.
MAT is also much faster than VITS, equally fast or faster than GRAD/GCFM at the same NFE, and only slightly slower than FS2 when at the fastest setting.

%estimated the intelligibility of the different conditions.

%\begin{itemize}
%\item Describe the objective aspects we evaluated, especially how we evaluated intelligibility
%\item Refer to appropriate table(s) for model size, memory use, and WER
%\item Write that our model was the second smallest (only Grad-TTS was smaller), but more importantly it used less memory during training than any other, which is the main bottleneck in training strong models
%\item Write that our unoptimised implementation achieved 2.3 updates per second during training, second only to Grad-TTS, which achieved 2.5 updates per second but also uses a more optimised implementation
%\item Describe how we calculated the synthesis speed for utterances of different lengths for the different synthetic conditions
%\item Refer to the appropriate scatterplot and discuss what we see (especially scaling behaviour)
%\item Refer to the table with the objective results (e.g., model size, synthesis speed), 
%\item Describe the setup of the listening test, similar to how we did it in our previous papers
%\item Refer to the results in a table, state which differences were significant, and discuss what we found and what it means
%\item Cite Ambika's SSW paper and the paper on range-equalising bias when commenting on the absolute numbers (which are consistent with the numerical range in our other evaluations of pre-trained systems, but lower than what is seen in papers about the same systems at machine-learning conferences)
%\end{itemize}

To evaluate the naturalness of the synthesised audio we ran a mean opinion score (MOS)
%\cite{itu1996telephone}
listening test.
We selected 40 utterances (4 groups of 10) of different lengths from the test set and synthesised each utterance using all conditions, loudness-normalising every stimulus
%to  $-$20 dB LUFS
using EBU R128.
%\cite{ebu2020loudness}.
80 subjects (self-reported as native English speakers using headphones) were crowdsourced through \href{https://prolific.co/}{Prolific} to listen to and rate these stimuli.
%80 self-reported native speakers of English (all of whom reported using headphones for the test) were crowdsourced through \href{https://prolific.co/}{Prolific} to listen to and rate these stimuli.
For each stimulus, listeners were asked ``How natural does the synthesised speech sound?'', and provided responses on an integer rating scale from 1 (``Completely unnatural'') to 5 (``Completely natural'') adopted from the Blizzard Challenge \cite{prahallad2013blizzard}.
Each group of 10 utterances was evaluated by 20 listeners, who were paid \textsterling 3 for a median completion time of 13 mins.
Inattentive listeners were filtered out and replaced in exactly the same way as in \cite{mehta2023overflow}.
In the end we obtained 800 ratings for each condition.
The resulting MOS values, along with confidence intervals based on a normal approximation, are listed in \cref{tab:results}.
We note that, since MOS values depend on many variables external to stimulus quality, e.g., listener demographics and instructions (see \cite{chiang23why, kirkland2023stuck}), they should not be treated as an absolute metric.
Comparing our MOS values to other papers is thus unlikely to be meaningful.%
%listener demographics and instructions; cf.\ \cite{chiang23why, kirkland2023stuck}.
%They should therefore not be treated as an absolute metric, and direct comparisons to scores other papers.
%We note that MOS values -- including ours -- are not an absolute metric, being substantially affected by differences in listener demographics and instructions; cf.\ \cite{chiang23why, kirkland2023stuck}.
%Direct comparison of scores across papers should therefore be avoided.%
%We note that the absolute numbers (and MOS values in general) are not comparable to those in other studies/papers, due to differences in listener demographics and instructions, as can be seen from the replication experiments in \cite{chiang23why, kirkland2023stuck}.%
\begin{figure}[!t]
\centering
\includegraphics[width=\columnwidth]{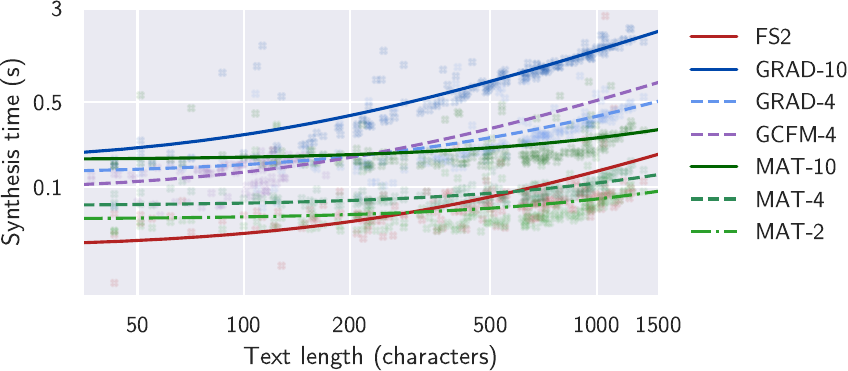}
\caption{Scatterplot of prompt length vs.\ synthesis time for acoustic models. Regression lines show as curves due to the log-log axes.}
\label{fig: rtf slopes}
\vspace{-\baselineskip}
\end{figure}

Applying $t$-tests to all pairs of conditions, all differences were found to be statistically significant at the $\alpha=0.05$ level except the pairs (MAT-10,MAT-4), (MAT-4,VITS), (VITS,MAT-2), (MAT-2,GCFM-4), and (GCFM-4,GRAD-10).
This means that MAT always had significantly better rated naturalness than GRAD for the same NFE, and always surpassed FS2.
Both the new architecture and training method contributed to the naturalness improvement, since MAT-4>GCFM-4>GRAD-4.
The fact that GRAD-10 was much better than GRAD-4 whilst MAT-10 and MAT-4 performed similarly suggests that GRAD requires many steps for good synthesis quality, whereas MAT reached a good level in fewer steps.
Finally, VITS performed similarly to MAT-2 and MAT-4 in terms of MOS.
MAT-10, although close to MAT-4 in rating, was significantly better than VITS.
For any given $n$, MAT-$n$ always scored higher than any system with equal or faster RTF.
In summary, Matcha-TTS achieved similar or better naturalness than all comparable baselines.
Finally, we evaluated how synthesis speed scaled with utterance length for the different models, by generating 180 sentences of different lengths using a GPT-2\footnote{\customurl{https://huggingface.co/gpt2}} model and plotting wall-clock synthesis time in \cref{fig: rtf slopes}, also fitting least-squares regression lines to the data.
The results show that MAT-2 synthesis speed becomes competitive with FS2 at longer utterances, with MAT-4 not far behind.
The major contributor to this appears to be the new architecture (since GRAD-4 and GCFM-4 both are much slower), and the gap from MAT to GRAD only grows with longer utterances.
%Interestingly, MAT-2 and MAT-4 outperform FS2 for longer utterances. Further exhibiting the practical benefits of Matcha-TTS.

\section{Conclusions and future work}
\label{sec:conclusion}
%\begin{itemize}
%\item Summarise what we did, similar to the abstract, but shorter (no background on the problem) and with more emphasis on our final outcome and its implications
%\item Describe one or two good directions for future research in terms of the model (e.g., probabilistic duration modelling) and its applications (e.g., spontaneous speech)
%\end{itemize}
We have introduced Matcha-TTS, a fast, probabilistic, and high-quality ODE-based TTS acoustic model trained using conditional flow matching.
The approach is non-autoregressive, memory efficient, and jointly learns to speak and align.
Compared to three strong pre-trained baselines, Matcha-TTS provides superior speech naturalness and can match the speed of the fastest model on long utterances.
Our experiments show that both the new architecture and the new training contribute to these improvements.
%The approach can match the speed of our fastest models
%that jointly learns to speak and align. Matcha-TTS employs optimal transport objectives for training resulting in high-quality speech synthesis with as low as 2 denoising iterations.

Compelling future work includes making the model multi-speaker, adding probabilistic duration modelling, and applications to challenging, diverse data such as spontaneous speech \cite{szekely2019spontaneous}.

% References should be produced using the bibtex program from suitable
% BiBTeX files (here: strings, refs, manuals). The IEEEbib.bst bibliography
% style file from IEEE produces unsorted bibliography list.
% 
% Gustav Eje Henter 2023-08-08: I changed this to a modern IEEEtran.bst
% This gives identical results but automatically abbreviates author names
% -------------------------------------------------------------------------
\bibliographystyle{IEEEtran}
\bibliography{refs_abbrev}

\end{document}